\documentclass[aps,showpacs]{revtex4}

\usepackage{amsmath} 
\usepackage{amssymb} 
\usepackage{epsfig}
\usepackage{xspace}

\newcommand{\p}{\partial} 
\newcommand{\tp}{\tilde{\psi}}
\newcommand{\bx}{{\bf x}}
\newcommand{\bv}{{\bf v}}
\newcommand{\bq}{{\bf q}}
\newcommand{\dd}{\text{\tiny $D$}}
\newcommand{\xx}{\text{\tiny $X$}}
\newcommand{\ew}{\text{\tiny EW}}
\newcommand{\kpz}{\text{\tiny KPZ}}
\newcommand{\tran}{\text{\scriptsize \rm tr}}
\newcommand{\pert}{\text{\scriptsize \rm p}}

\newcommand{\tps}{\tilde{\partial_s}}
\newcommand{\gk}{\Gamma_k} 
\newcommand{\tih}{\tilde{h}}
\newcommand{\tj}{\tilde{j}}
\newcommand{\anz}{{\em Ansatz}\xspace}
\newcommand{\nequ}{non-equilibrium\xspace}
\newcommand{\npt}{non-perturbative\xspace}
\def\nbR{\ensuremath{\mathrm{I\!R}}} 
\newcommand{\ptm}{\text{-}}

\begin{document}
\title{Strong-Coupling Fixed Point of  the KPZ  Equation}

\author{L\'eonie Canet}

\affiliation{                    
   Service de Physique de l'\'Etat Condens\'e, Orme des Merisiers -- CEA Saclay, 91191 Gif-sur-Yvette, France}



\begin{abstract}
{\em NOTE: This paper presented the first attempt to tackle the Kardar-Parisi-Zhang (KPZ) equation using \npt renormalisation group  (NPRG) methods. It exploited the most natural and frequently used approximation scheme within the NPRG framework, namely the derivative expansion (DE).  However, the latter approximation turned out to yield  unphysical critical exponents in dimensions  $d\ge 2$ and, furthermore, hinted at very poor convergence properties of the DE. The author has since realized that in fact, this approximation may not be valid for the KPZ problem, because of  the very nature of the KPZ interaction, which is not {\em potential} but {\em derivative}.  The probable failure of the DE is a very unusual -- and instructive -- feature within the NPRG framework. As such, the original work, unpublished, is left available on the arXiv and can be found below. 

Added note:  the key to deal with the KPZ problem using NPRG lies in not truncating the momentum dependence of the correlation functions, which is investigated in a recent work {\em arXiv:0905.1025}.} \\

We present a new approach to the Kardar-Parisi-Zhang (KPZ) equation based on the \npt renormalisation group (NPRG). The NPRG flow equations derived here, at the lowest order of the derivative expansion, provide a stable strong-coupling fixed point in all dimensions $d$, embedding in particular the exact results in $d=0$ and $d=1$.  However, it yields at this order unreliable  dynamical and roughness exponents $z$ and $\chi$ in higher dimensions, which suggests that a richer approximation is needed to investigate the property of the rough phase in $d \ge 2$.
\end{abstract}


\maketitle

The  Kardar-Parisi-Zhang (KPZ) equation, though originally introduced as a coarse-grained description of \nequ interface growth \cite{kardar86}, has acquired a broader significance over the past decades as a simple model for  generic scale invariance and \nequ phase transitions \cite{halpin95}. It is indeed intimately related to many other important physical systems, such as randomly stirred fluid (Burgers equation) \cite{forster77},
  directed polymers in random media \cite{kardar87},  dissipative transport \cite{beijeren85,janssen86} or magnetic flux lines in superconductors \cite{hwa92}. The KPZ equation has thus emerged as one of the fundamental theoretical models to investigate universality classes in \nequ scaling phenomena and phase transitions \cite{halpin95}.

 It is a non-linear Langevin equation which describes the large-distance, long-time dynamics of the growth process specified by a single-valued height function $h(\bx,t)$ on  a $d$-dimensional substrate $\bx \in \nbR^d$:
\begin{equation}
\p_t h(\bx,t) = \nu\,\nabla^2 h(\bx,t) \, + \,\lambda/2\,\big(\nabla h(\bx,t)\big)^2 \,+\,\eta(\bx,t),
\label{eqkpz}
\end{equation}
where $\eta(\bx,t)$ is a zero mean uncorrelated noise with variance  
$\langle \eta(\bx,t)\eta(\bx',t')\rangle = 2 D \delta^d(\bx-\bx')\,\delta(t-t')$. This equation reflects the competition between the surface tension smoothing force $\nu\nabla^2 h$, the preferential growth along the local normal to the surface represented by the non-linear term and the Langevin noise $\eta$ which tends to roughen the interface and mimics the stochastic nature of the growth. 

The objective is to determine the profile of the stationary interface, characterised by the two-point correlation function $C(|\bx-\bx'|,t-t') \equiv \langle [h(\bx,t) - h(\bx',t')]^2\rangle$ and, in particular,  its large-scale  properties where  $C$ is expected to assume the scaling form
$C(L,\tau) = L^{2\chi}\,f({\tau}/{L^z})$, where $\chi$ and $z$ are the roughness and  dynamical exponents respectively. These two exponents are not independent since the Galilean symmetry \cite{forster77} --- the invariance of Eq. (\ref{eqkpz}) under an infinitesimal tilting of the interface --- enforces the scaling relation $z+\chi = 2$.

While the KPZ equation is solvable in $d=1$ \cite{kardar87,hwa91} and on a Bethe lattice \cite{imbrie88}, its complete theoretical understanding  is still lacking. The one-dimensional case is special due to the incidental existence of a fluctuation-dissipation theorem
 (FDT) which yields the exact result $\chi=1/2$ and $z=3/2$. 
For $d>2$, Eq. (\ref{eqkpz}) entails two different regimes, separated by a critical value  $\lambda_c$ of the non-linear coefficient \cite{kardar86,forster77}. In the weak-coupling regime ($\lambda < \lambda_c$),  the behaviour is governed by the $\lambda= 0$ fixed point ---  corresponding to the linear  Edwards-Wilkinson equation \cite{halpin95} --- with exponents $\chi = (2-d)/d$ and $z=2$. In the strong-coupling regime ($\lambda > \lambda_c$), the non-linearity becomes relevant and despite considerable efforts, the statistical properties of the rough interface above $d=2$ remain very controversial.

 Indeed, few analytical approaches have proved successful. While the  numerical integration of discretised versions of Eq. (\ref{eqkpz}) is plagued with singularities \cite{newman96}, perturbative  RG, even re-summed to all orders, 
 fails to describe the strong-coupling phase \cite{wiese97}.
 On the other hand, surprisingly accurate exponents can be worked out from mode-coupling theory  \cite{beijeren85,bouchaud93}  but the latter would deserve to be put on firmer grounds since it remains a `ad-hoc' approximation \cite{hwa91}. 
As a consequence, the very existence of an upper critical dimension $d_c$ for this problem is still much debated.
Most of the analytical arguments \cite{bouchaud93,halpin89,wiese97} suggest $d_c \simeq 4$, whereas numerical simulations and real space calculations  find no hints for a finite $d_c$ \cite{tang92}. 

In this context, a controlled  analytical approach appears highly desirable. This  motivates our  resort to the NPRG method \cite{berges02,bagnuls01}.
 This formalism, which has allowed to tackle notoriously difficult strong-coupling problems at equilibrium \cite{berges02}, has been recently extended to \nequ systems \cite{canet04a,canet04b}. It has given rise to important progress for reaction-diffusion processes, revealing  crucial \npt effects \cite{canet04a,canet04b}. It therefore stands as a promising candidate to investigate the KPZ problem and we show in this work that it indeed provides the strong-coupling fixed point  that allows to describe the rough phase in $d\geq 2$.\\

Our starting point is the field theory corresponding to Eq. (\ref{eqkpz}), with response functional:
\begin{equation}
{\cal S}[h,\tih] = \int d^d \bx\,dt \, \left\{ \tih\left[\p_t h -\nu \,\nabla^2 h - \frac{\lambda}{2}\,(\nabla h)^2 \right] - D\, \tih^2  \right\},
\label{jdd}
\end{equation}
 which follows from the Janssen - De Dominicis formalism \cite{janssen76}. 
 This field theory can  be investigated using the NPRG.
 We don't detail here its implementation but rather set out its principle (see \cite{berges02} for reviews).
This formalism relies on Wilson's RG idea, which consists in building  a sequence of scale-dependent effective models,
that interpolate smoothly between the short-distance physics at the
(microscopic) scale $k=\Lambda$ and the long-distance one at $k=0$,
 through progressively averaging over fluctuations.
Rather than expressing --- as in the original Wilsonian formulation --- the flow of effective Hamiltonians for the long-distance modes,
one can work out the flow of effective `free energies' $\gk$ for the short-distance modes that have been integrated, following \cite{berges02}. Thus, at $k=\Lambda$, no  fluctuation has yet been included
and $\Gamma_{\Lambda}$ coincides with the microscopic action ${\cal S}$, while at  $k=0$, all  fluctuations have been integrated out and $\Gamma_{0}$ is the analogue of the Gibbs free energy $\Gamma$ at thermal equilibrium, in that it encompasses the large-scale (and long-time) properties of the system.
To construct $\gk$, one needs to suppress the low-momentum modes. This is achieved by adding a scale-dependent   term to the original action \cite{berges02,canet04c}:
\begin{equation}
\Delta {\cal S}_k[h,\tih] = \int_{\bq,\omega} \bq^2 \,\tih(\ptm \bq,\ptm \omega) \,R^\nu_k(\bq^2)\,h(\bq,\omega) - 2\,\tih(\ptm \bq,\ptm \omega) \,R^\dd_k(\bq^2)\,\tih(\bq,\omega) 
\label{dsk}
\end{equation}
where the cutoff function $R^\xx_k(\bq^2) = X_k\, r(\bq^2/k^2)$ is large for small momenta $|\bq| \leq k$  so that the low-momentum modes are decoupled  and  vanishes for large momenta $|\bq| \geq k$.
The generating
 functionals ${\cal Z}_k[j,\tj]= \int {\cal D}h\, {\cal D}i \tih\, \exp(- {\cal S}- \Delta {\cal S}_k +  \int j h +  \int \tj \tih)$ become therefore
$k$-dependent.
$\gk$ is the (modified)  Legendre transform of   
$\log {\cal Z}_k[j,\tj]$:
$ \Gamma_k[\psi,\tp] =- \log {\cal Z}_k[j,\tj] + 
\int j \psi +\int \tj \tp -\Delta {\cal S}_k[\psi,\tp]$,
 where the last term ensures
that $\Gamma_k$ has the proper limit at $k=\Lambda$: $\Gamma_{k=\Lambda}\sim
{\cal S}$ \cite{berges02,canet04c}. $\gk$ is a functional of the conjugate fields
$\psi=\delta \log {\cal Z}_k/\delta j$ and $\tp=\delta \log {\cal Z}_k/\delta
\tj$. 
An exact functional differential equation governs
 its RG flow  under an infinitesimal change of
the scale $s=\log(k/\Lambda)$
\cite{berges02,canet04c}:
\begin{equation}
\partial_s \Gamma_k = \frac{1}{2} \tps {\rm Tr} \int_{q,\omega} \ln \hat P[\psi,\tp],
\label{dkgam}
\end{equation}
where $\tps \equiv \p_s R^\dd_k\, \p_{R^\dd} + \p_s R^\nu_k \,\p_{R^\nu}$ and $\hat P[\psi,\tp]$ is the $2\times
2$ matrix of second derivatives of $(\Gamma_k$ + $\Delta {\cal S}_k)$ with respect to (w.r.t.) $\psi$ and
$\tp$. $\hat{P}^{-1}$ embodies the full ( {\it i.e.} functional) renormalised propagator of the theory. 
Obviously, Eq.~(\ref{dkgam}) cannot be solved exactly
and one usually truncates it.  However, as the truncations used do not rely on the smallness of a parameter, the approach remains in essence \npt. 

Since the critical physics corresponds to the long-distance, long-time limit,
a sensible truncation  consists  in
expanding  $\Gamma_k$ in powers of  gradients \cite{berges02} and time derivatives.
At leading order in derivatives, the  \anz for $\Gamma_k$ related to the field theory (\ref{jdd})  writes:
\begin{equation}
\gk=\int d^d\bx \, d t\, \Big\{\tp\,\Big[  \mu_k\,\p_t - \nu_k\, \nabla^2\Big]\psi
 -\frac{1}{2} \lambda_k\, \tp\,\big(\nabla \psi\big)^2 - D_k\, \tp^2.
\label{anz}
\end{equation}
This \anz must be  invariant under 
 a Galilean transformation, which writes:
\begin{equation}
\Big\{
\begin{array}{l c l}
\psi(\bx,t) & \to & \psi(\bx +\lambda_k \bv t,t) + \bv.\bx \\
\tp(\bx,t) & \to & \tp(\bx+\lambda_k \bv t,t).
\end{array}
\label{galsym}
\end{equation}
Moreover,  the special time reversal symmetry that yields FDT in $d=1$ 
 can be encoded as an invariance  under the field transformation 
\begin{equation}
\Big\{
\begin{array}{l c l}
\psi(\bx,t) & \to & -\psi(\bx,-t) \\
\tp(\bx,t) & \to & \tp(\bx,-t) +\frac{\nu_k}{2 D_k} \Delta\psi(-t)
\end{array}
\label{fdtsym}
\end{equation}
which is satisfied both by the bare action (\ref{jdd}) and by our \anz (\ref{anz}). 

First, note that  $\Delta {\cal S}_k[\psi,\tp]$ defined by (\ref{dsk}) has been  deviced to respect both the  Galilean and time reversal invariance.
 Then, the requirement that (\ref{anz}) is Galilean-symmetric implies $\mu_k\,\lambda_k = \lambda_k$, that is  $\mu_k=1$, which means that fields are not renormalised --- they acquire no 
anomalous dimension. 
 This feature can be checked explicitly 
by calculating the flow equation for $\mu_k$ --- one finds $\p_s \mu_k = 0$.
  Also, $\lambda_k$,  structure constant of the Galilean symmetry group, should not be renormalised (as can be inferred from the
 corresponding Ward identities \cite{frey94}). We show below that this is indeed automatically satisfied in our formalism. 

 The \npt flow equations associated with the \anz (\ref{anz}) can be derived as follows. The coupling constants
 $\nu_k$, $\lambda_k$ and $D_k$  are defined from (\ref{anz})
 by:
\begin{equation}
 \nu_k = \p_{q^2} \frac{\p^2 \gk}{\p\psi \,\p\tp},\hbox{\hspace{1cm}}  \lambda_k =  \p_{q^2} \frac{1}{\tp}\frac{\p^2 \gk}{\p\psi^2},\hbox{\hspace{0.5cm}and\hspace{0.5cm}}  D_k =  -\frac{1}{2}\frac{\p^2 \gk}{\p\tp^2}
\end{equation}
 evaluated  at a uniform and stationary field configuration  $(\psi_0,\tp_0)$.
 Their flow equations follow from  Eq. (\ref{dkgam}) by taking the corresponding derivatives  w.r.t.  fields (and momentum), and evaluating them at $(\psi_0,\tp_0)$. After integrating over the frequency $\omega$, we obtain: 
\begin{equation}
\p_s D_k = \frac{\lambda_k^2}{4} \tps \int \frac{d^d \bq}{(2\pi)^d} \,\frac{\bq^4\, (D_k + R^\dd_k)^2}{\Big[\bq^4\,(\nu_k + R^\nu_k)^2 -2\, \bq^2\,\lambda_k\, \tp_0 \,(D_k + R^D_k )\Big]^{3/2}},
\label{dtD}
\end{equation}
\begin{equation}
\p_s \nu_k = \frac{\lambda_k^2}{4\,d} \tps \int \frac{d^d \bq}{(2\pi)^d} \,\frac{\bq^4\,(2-d)\, (D_k + R^\dd_k)\, (\nu_k + R^\nu_k)+ 2\, \bq^6 \Big((D_k + R^\dd_k)\,\p_{\bq^2}R^\nu_k - (\nu_k + R^\nu_k)\,\p_{\bq^2}R^\dd_k\Big)}{\Big[\bq^4\,(\nu_k + R^\nu_k)^2 -2\,\bq^2\, \lambda_k\, \tp_0\,(D_k + R^\dd_k )\Big]^{3/2}}
\label{dtnu}
\end{equation}
\begin{equation}
\hbox{and\hspace{2cm}}  \p_s \lambda_k = 0.
\end{equation}
 The critical regime corresponds to a fixed point (FP) of the (dimensionless) flow of $\gk$, where the coefficients $D_k$ and $\nu_k$ --- denoted $X_k$ ---  assume  a power-law behaviour $X_k \sim k^{-x_{\xx}}$ with
 $x_\xx\equiv  -\p_s \ln X_k$. Since, at criticality, $C\sim k^{-2\chi}$ and $\omega\sim k^z$, the anomalous dimensions $x_\nu$ and $x_\dd$ (at the FP) are related to the critical exponents $\chi$ and $z$ through:
 $z = 2 - x_\nu$ and $\chi=1-d/2 + (x_\dd - x_\nu)/2$.
 We emphasize that $\p_s \lambda_k = 0$ is verified at all scales $k$ and for all $\tp_0$. This implies that the Galilean invariance is preserved all along the flow at any non-zero FP, which  in turn enforces the relation  $x_\dd -3\, x_\nu = d- 2$ (or equivalently $z+\chi =2$). Moreover, one can check  from Eqs. (\ref{dtD}) and (\ref{dtnu}) that, upon setting $D_k=\nu_k$ and $d=1$, one obtains $\p_s D_k = \p_s \nu_k$ at all scales $k$, {\it i.e.} FDT is satisfied all along the flow in $d=1$.
\\

One usually considers the combination $\bar{\lambda}_k \equiv \lambda_k^2\, D_k/\nu_k^3$. 
 Upon introducing the dimensionless variables $\phi = (k^{d+2} \,\nu_k/D_k)^{-1/2}\,\tp_0$, $g_k = k^{d-2}\, \bar{\lambda}_k$, $y=\bq^2/k^2$ and the explicit cut-off function $r(y) = (1/y - 1)\,\theta(1-y)$ \cite{litim01} --- where $\theta$ denotes the Heaviside step function \footnote{This cut-off function has been chosen because it enables one to perform analytically the integration over momentum.} --- we obtain:
\begin{equation}
\p_s  \ln D_k =\frac{-4 \,g_k \, v_d}{d \,(2 + d) \,m^{5/2}}\Big[2+d  + 2\,x_\dd -3 \, x_\nu 
+\phi  \,\sqrt{g_k}\Big(d+2- x_\dd\Big) \Big]\nonumber
\end{equation}
\begin{equation}
\p_s \ln \nu_k =\frac{4 \,g_k \, v_d}{d^2 \,(2 + d) \,m^{5/2}}\Big[d^2 - 4 -2\,x_\dd +(4-d)\, x_\nu 
+\phi  \,\sqrt{g_k}\Big(d^2 - 4 +(3\,d-2)\,x_\dd -4(d-1)\, x_\nu \Big) \Big] \nonumber
\end{equation}
\begin{equation}
\p_s  g_k = (d-2)\, g_k + g_k\,\Big(\p_s  \ln D_k -3 \,\p_s \ln \nu_k \Big)
\label{dtg}
\end{equation}
where $m\equiv 1-2\, \phi\,\sqrt{g_k}$ and $v_d^{-1}=2^{d+1}\,\pi^{d/2}\,\Gamma(d/2)$.\\

\begin{figure}[h]
\centering\includegraphics[height=86mm,angle=-90]{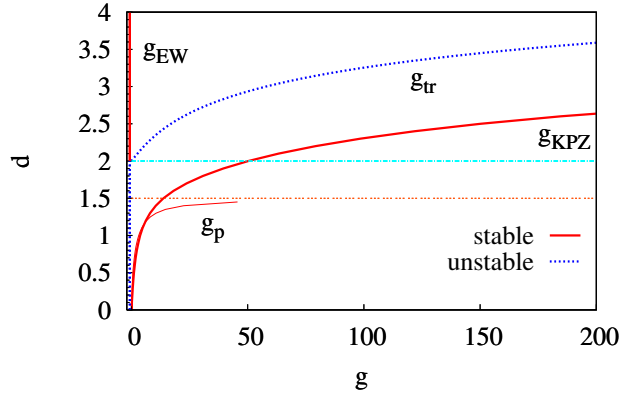}
\caption{Phase diagram of the KPZ problem. $g_{\ew}$, $g_{\tran}$, $g_{\pert}$ and $g_{\kpz}$ stands for
 Edwards-Wilkinson, transition, perturbative and strong-coupling FP respectively, see text for comments. This diagram corresponds to $\phi=0$ but remains qualitatively unchanged for non-vanishing $\phi$. The horizontal dashed lines are guidelines for the eyes.}
\label{diag}
\end{figure}

  These equations are valid for arbitrary $d$ and $g_k$ and, as we show below, they  provide, at a qualitative level, the {\it complete} phase diagram of the KPZ equation in the $(g,d)$ plane, which is displayed on Fig. \ref{diag}. This constitutes  the main result of this work:  the NPRG equations derived here 
entails a stable strong-coupling FP that corresponds to the rough phase {\it in all dimensions}, yielding in particular the exact results in $d=0$ and $d=1$.
%
%

We now present in details the FP solutions  of  the NPRG flow equations (\ref{dtg}).  For $\phi=0$, they can be solved analytically and
  exhibit three FP, denoted $g_{\ew}$,  $g_{\tran}$  and  $g_{\kpz}$. 
 The FP $g_{\ew}=0$ corresponds to the linear Edwards-Wilkinson FP characterised by $x_\nu = x_\dd =0$, ({\it i.e.} $z=2$, $\chi=1-d/2$). 
 The FP $g_{\tran}$ and $g_{\kpz}$ are both Galilean-symmetric, {\it i.e.} satisfy $z+\chi=2$.  The FP $g_{\tran}$ is unstable and hence drives a transition for $d>2$. We find $x_\nu \lesssim 0$, departing all the more from zero than the dimension increases, which indicates that the accuracy of the approximation deteriorates as $d$ grows. This result means that the dynamical exponent reaches the upper bound $z=2$ at the transition,
 which is consistent with the exact result $\chi=0$  ensuing from the perturbative RG re-summed to all orders \cite{wiese97}.  We now focus on  $g_{\kpz}$, which is a stable FP in all $d$.
 First, it yields in $d=1$ the exact exponent $z=3/2$. This result follows from the time reversal  invariance (\ref{fdtsym}) of the flow equations in $d=1$.
   The zero-dimensional exact result $z=4/3$ is also recovered. 
 Moreover, $g_{\kpz}$ can be followed in all dimensions and allows to investigate the properties of the high-dimensional rough phase. 
 This shows that our method is  genuinely nonperturbative.
 Indeed, conversely,  the perturbative one-loop RG FP $g_{\pert}$ \cite{medina89,frey94} shown on Fig. \ref{diag} --- which  also simply stems from the ${\cal O}(g_k^2)$ expansion of Eqs. (\ref{dtg}) --- diverges in $d=3/2$ where the perturbative scheme breaks down. Moreover the re-summed to all order perturbative RG flow equation for $g$ does not yield any non-trivial stable FP, {\it i.e.} it cannot access the strong-coupling regime \cite{wiese97}.\\

 Unfortunately, while the \anz (\ref{anz}) is sufficient to provide the whole phase diagram at a qualitative  level, it does not yield correct critical exponents in $d\ge 2$: in fact the value of $z$ starts decreasing for $d \ge2$, which is unphysical.

Let us now discuss possible improvements.
 A key advantage of the NPRG formalism  is that it can be improved in a systematic way.
 Indeed, efficient \npt approximation schemes have been developed and thoroughly studied  during the last decade \cite{berges02}. They have shown that the lowest order truncation (analogous to the \anz used here) always turns out to be sufficient to  capture in a robust way all the qualitative features of a model \cite{berges02,canet04a,canet04b}. 
 For most models, even out of equilibrium (reaction-diffusion  processes for instance), going to the next order in the derivative expansion becomes increasingly tedious but allows to obtain very accurate exponents (and other physical quantities) \cite{berges02,canet03b}. 

\begin{figure}[h]
\centering\includegraphics[height=94mm,width = 78mm,angle=-90]{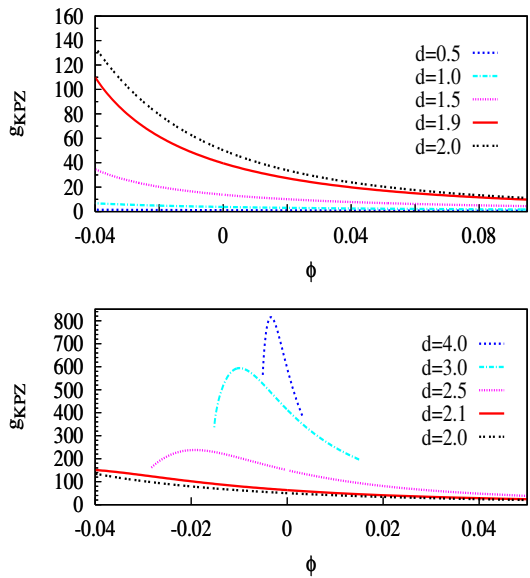}
\caption{Strong-coupling FP $g_{\kpz}$ as a function of the dimensionless response field $\phi$. The curves below (top row) and above (bottom row) $d=2$ are split in two sets for clarity. Note that in $d=4$, $g$ has been rescaled to $g/5$ to fit in.}
\label{expo}
\end{figure}

 For the KPZ action (\ref{anz}), the next approximation level within the  derivative expansion  consists in considering field-dependent functions  $\nu_k(\tp)$ and $D_k(\tp)$ (there is no $\psi$ dependence at this order due to the symmetries). We have investigated this  level of truncation. Prior to discussing the results, let us state the outcome:
 unfortunately, while confirming the {\it qualitative} phase diagram found at the lowest order, it  remains unsufficient to obtain meaningfull exponents for $d\ge 2$.
This failure seems  to root in very slow convergence properties of the derivative expansion for $d\ge2$, as can be  illustrated by probing the influence of  the dimensionless response field value $\phi$ --- which is related to the noise strength. The strong-coupling FP solution of Eqs. (\ref{dtg}) can be determined numerically  for different finite $\phi$, which has been achieved.  The dependence of $g_{kpz}$ on  $\phi$ in various dimensions is displayed on Fig. \ref{expo}.
 It shows that the behaviour of $g_\kpz$ seems to undergo a noticeable change crossing  $d=2$. 
%
Below two dimensions, the FP coordinate $g_{\kpz}(\phi)$ shows  a monotonic and weak dependence on the field $\phi$, whereas above $d=2$ it acquires a non-trivial shape, the two regimes appearing separated by the $d=2$ curve (see Fig. \ref{expo}, note the different  scales for $g$).  This  suggests that the convergence of the scheme deteriorates above $d=2$. In fact, the specificity of the KPZ model, compared with other non-equilibrium models such as reaction-diffusion processes for instance, is the 
 nature of its interaction, which is derivative, and not potential.
 Whereas the derivative expansion  converges very rapidly for models with potential interactions \cite{berges02,canet04a,canet04b}, it would require for the KPZ model many orders to lead to quantitative exponents, which is not conceivable. This suggests that the derivative expansion is not the best strategy to investigate the physics of the rough phase. One should device an alternative scheme of approximation, adapted to derivative interactions.
 This   will be presented elsewhere  \cite{canet06}.\\

To summarize,  we presented in this work the first theoretical  approach to the KPZ problem, which entails the strong-coupling FP describing the rough phase in all dimensions, embedding in particular  the exact results in $d=0$ and 1. While the approximation scheme used here, the derivative expansion, provides a consistent and robust picture of the whole phase diagram
 at a {\it qualitative level},  it is not sufficient to obtain physical critical exponents above $d=2$. An alternative approximation scheme is needed for a  thorough analysis of the properties of the strong-coupling regime.\\

\begin{acknowledgments}
The author is particularly indebted to J.-P. Bouchaud, B. Delamotte, H. Chat\'e and M.A. Moore for fruitful discussions and wishes to thank B. Delamotte and H. Chat\'e for careful readings of the manuscript.  Part of this  work benefits from the financial support by the European Community's Human Potential Programme under contract HPRN-CT-2002-00307, DYGLAGEMEM.

\end{acknowledgments}


\begin{thebibliography}{10}

\bibitem{kardar86}
M. Kardar, G. Parisi, and Y.-C. Zhang, Phys. Rev. Lett. {\bf 56},  889  (1986).

\bibitem{halpin95}
T. Halpin-Healy and Y. Zhang, Phys. Rep. {\bf 245},  218  (1995).

\bibitem{forster77}
D. Forster, D.~R. Nelson, and M.~J. Stephen, Phys. Rev. A {\bf 16},  732
  (1977).

\bibitem{kardar87}
M. Kardar, Nucl. Phys. B {\bf 290},  582  (1987).

\bibitem{beijeren85}
H. van Beijeren, R. Kutner, and H. Spohn, Phys. Rev. Lett. {\bf 54},  2026
  (1985).

\bibitem{janssen86}
H.~K. Janssen and B. Schmittmann, Z. Phys. B {\bf 63},  517  (1986).

\bibitem{hwa92}
T. Hwa, Phys. Rev. Lett. {\bf 69},  1552  (1992).

\bibitem{hwa91}
T. Hwa and E. Frey, Phys. Rev. A {\bf 44},  R7873  (1991).

\bibitem{imbrie88}
J. Imbrie and T. Spencer, J. Stat. Phys. {\bf 52},  609  (1988).

\bibitem{newman96}
T.~J. Newman and A.~J. Bray, J. Phys. A {\bf 29},  7917  (1996).

\bibitem{wiese97}
K.~J. Wiese, Phys. Rev. E {\bf 56},  5013  (1997).

\bibitem{bouchaud93}
J.-P. Bouchaud and M.~E. Cates, Phys. Rev. E {\bf 47},  R1455  (1993).

\bibitem{halpin89}
T. Halpin-Healy, Phys. Rev. Lett. {\bf 62},  442  (1989).

\bibitem{tang92}
L.-H. Tang, B.~M. Forrest, and D.~E. Wolf, Phys. Rev. A {\bf 45},  7162
  (1992).

\bibitem{berges02}
J. Berges, N. Tetradis, and C. Wetterich, Phys. Rep. {\bf 363},  223  (2002).

\bibitem{bagnuls01}
C. Bagnuls and C. Bervillier, Phys. Rep. {\bf 348},  91  (2001).

\bibitem{canet04a}
L. Canet, B. Delamotte, O. Deloubri\`ere, and N. Wschebor, Phys. Rev. Lett.
  {\bf 92},  195703  (2004).

\bibitem{canet04b}
L. Canet, H. Chat\'e, and B. Delamotte, Phys. Rev. Lett. {\bf 92},  255703
  (2004).

\bibitem{janssen76}
H.~K. Janssen, Z. Phys. B {\bf 23},  377  (1976).

\bibitem{canet04c}
L. Canet, Phys. Ann. Fr. {\bf 29},    (2004).

\bibitem{frey94}
E. Frey and U.~C. T\"auber, Phys. Rev. E {\bf 50},  1024  (1994).

\bibitem{litim01}
D.~F. Litim, Phys. Rev. D {\bf 64},  105007  (2001).

\bibitem{medina89}
E. Medina, T. Hwa, M. Kardar, and Y.-C. Zhang, Phys. Rev. A {\bf 39},  3053
  (1989).

\bibitem{canet03b}
L. Canet, B. Delamotte, D. Mouhanna, and J. Vidal, Phys. Rev. B {\bf 68},
  064421  (2003).

\bibitem{canet06}
L. Canet, H. Chat\'e, B. Delamotte, and N. Wschebor, arXiv:0904.1025.

\end{thebibliography}

\end{document}